\documentclass[11pt]{article}
\usepackage{moriond,epsfig,wrapfig}
\def\Journal#1#2#3#4{{#1} {\bf #2}, #3 (#4)}

\def\NIMA{{\em Nucl. Instrum. Methods} A}

\def\PRL{\em Phys. Rev. Lett.}
\def\PRD{{\em Phys. Rev.} D}

\def\be{\begin{equation}}
\def\ee{\end{equation}}
\def\bea{\begin{eqnarray}}
\def\eea{\end{eqnarray}}

\begin{document}
\vspace*{4cm}
\title{THE UNDERLYING EVENT IN HADRON-HADRON COLLISIONS}
\author{V. TANO}
\address{Max-Planck-Institut f\"ur Physik,\\
F\"ohringer Ring 6, Munich, Germany}
\maketitle\abstracts{
We measured the ambient energy in jet events using data of the CDF experiment~\cite{CDF} 
at center-of-mass energies of  1800 and 630 GeV by examining the 
transverse momentum of charged particles in a region of the detector far away from the two most energetic jets in the event. The energy is compared to that found in minimum bias events. CDF data are compared to Monte Carlo predictions: neither Herwig~\cite{Herwig} nor Pythia~\cite{Pythia} can reproduce the data in  detail.
}

\section{Introduction}

In hadron-hadron collisions, besides the interaction between the two partons responsible for the hard scattering, there may be an additional interaction between the beam remnant partons. Usually, this interaction is {\it soft}, involving low momentum transfer, therefore perturbative QCD can not be applied and it has to be described by models. Contributions to the final energy may come from additional gluon radiation from either the initial-state or final-state partons, as well as additional semi-hard interactions (double parton scattering) from the initial-state constituents. This energy, which does not depend on the two partons responsible for the hard scattering, defines the so called {\it underlying event}.\par
At CDF, jet clustering is based on a fixed cone algorithm. In order to compare jet data to perturbative QCD calculations, the underlying event energy has to be estimated and subtracted from the jet energy~\cite{jet_inc_sigma}.  
At a center-of-mass energy of 1800 GeV, the uncertainty in the amount of energy to be subtracted is the largest experimental error for low E$_t$ jets ($\sim 30\%$). 
The actual assumption is that the underlying event energy is similar to the energy observed  in minimum bias events. The minimum bias trigger selects the majority of the inelastic non-diffractive cross section. We examine both jet and minimum bias events in order to study this assumption.

\section{Data sample}
At a center-of-mass energy of 1800 GeV, CDF collected jets using four different triggers with E$_t$ of the leading jet larger than 100, 70, 50, 20 GeV. We consider only events with the leading jet in the central detector region ($0.1<|\eta|<0.7$) and with E$_t$ of the leading jet larger than 130, 100, 75, 40 GeV respectively for the four jet samples. At 630 GeV, two triggers were used to collect data, requiring the E$_t$ of the leading jet to be larger than 5 and 15 GeV. Again, we select only events with the leading jet in the central pseudorapidity region and with E$_t$ of the leading jet larger than 20 and 30 GeV respectively. \par 
The minimum bias trigger requires a coincidence in the East and West beam-beam counters, two scintillation counters located along the beam line at $\pm$ 5.8 m from the interaction point and covering the pseudorapity region: $3.24<|\eta|<5.90$.\par
At small energies, calorimeter studies depend too strongly on details of the  detector simulation, therefore we consider only tracks from charged particles. The main tracking device in CDF is the CTC, a cylindrical drift chamber, which extend up to $|\eta| = \pm2.4$ and provides a precise momentum measurement in the central pseudorapidity region $|\eta|<1$ with a momentum resolution better than $\delta p_t/p_t{^2}\leq 0.002$ (GeV/c)$^{-1}$.

\section{Results}

\begin{wrapfigure}[23]{r}{4cm}
\psfig{file=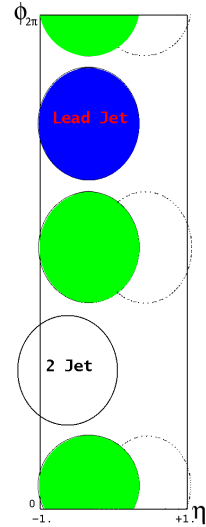,height=7.cm,clip=}
\caption{{\label{Fig-phase-space}} Example of two jet events  in the detector region under study. Cones used for the determination of underlying event contribution are at $\eta = \eta_{LeadJet}$ and $\phi = \phi_{LeadJet} \pm 90^{\circ}$.}
\end{wrapfigure}

In jet events, we examine the transverse momentum of the charged particles in the central rapidity region inside two cones of radius $0.7$ at the same rapidity and at $\pm 90^\circ$ in azimuth from the most energetic jet in the event. The cone size is the same used to reconstruct jets in the inclusive jet cross section analysis at CDF.
In Figure~\ref{Fig-phase-space} the central detector region is shown  \it{unrolled}\rm: $\eta$ ranges are between  -1 and +1, while $\phi$ goes from $0^{\circ}$ to $360^{\circ}$. 
The leading jet cone  and the two cones under study are shown. 
The two cones are used to study the underlying event energy because they are supposed to be in a semi-quiet region, far away from the two leading jets, but still in the central rapidity region. Given the non-uniform response of the CDF detector as a function of rapidity, the latter criterion is essential.
For each event we label the cone which has the maximum momentum ($max$ cone) and the cone with minimum momentum ($min$ cone). This is useful because NLO perturbative corrections to the $2\rightarrow2$ hard scattering can contribute only to one of these two regions ~\cite{marchesini_webber}. The difference between the $max$ and the $min$ cone provides information on this contribution, while the $min$ cone  gives an  indication of the
level of the underlying event from non-NLO sources. \par
\begin{figure}[h]
\centering
\mbox{
\hspace{-.5cm}
\includegraphics[height=7.8cm]{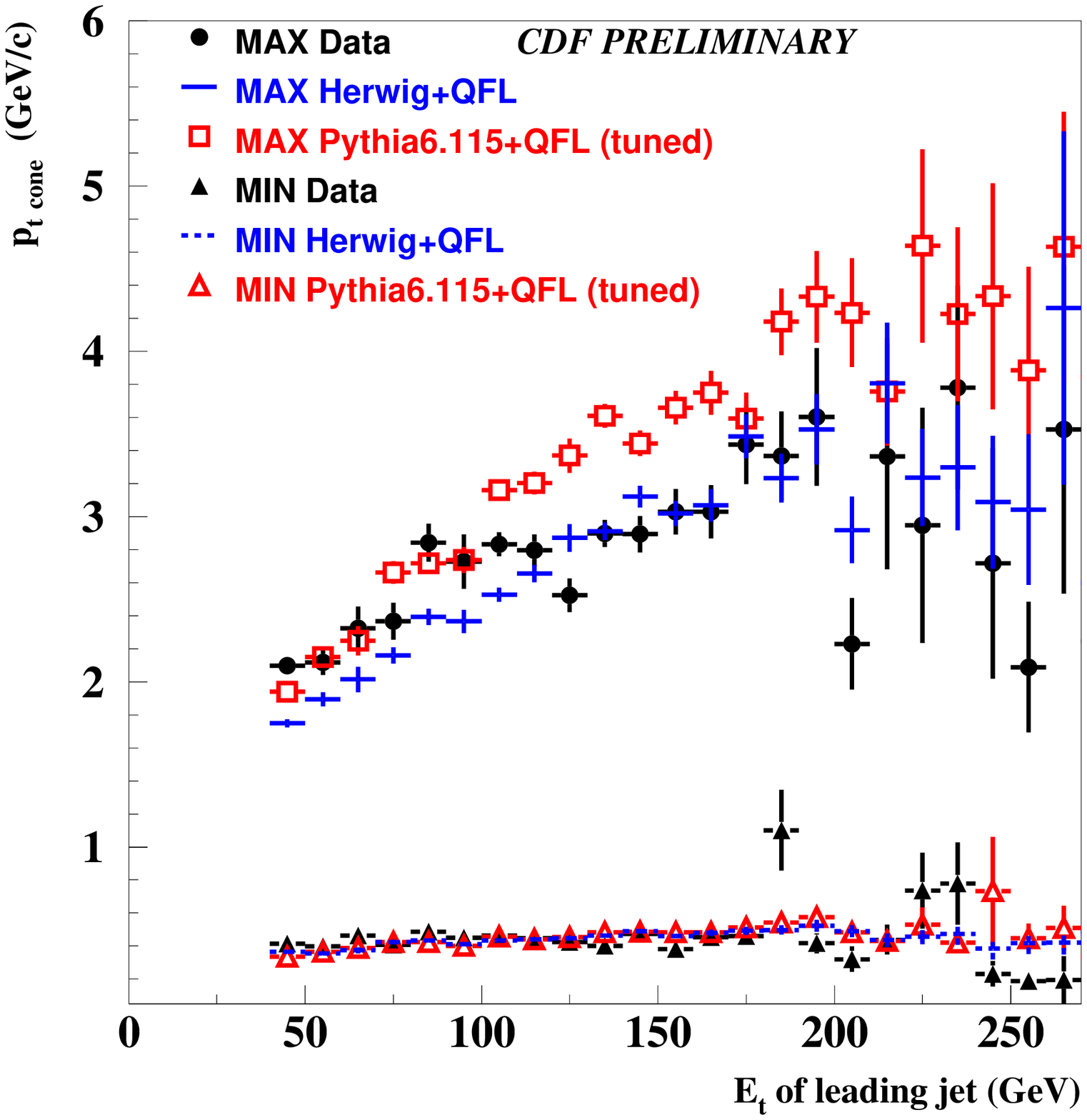}
\hspace{-0.5cm}
\includegraphics[height=7.8cm]{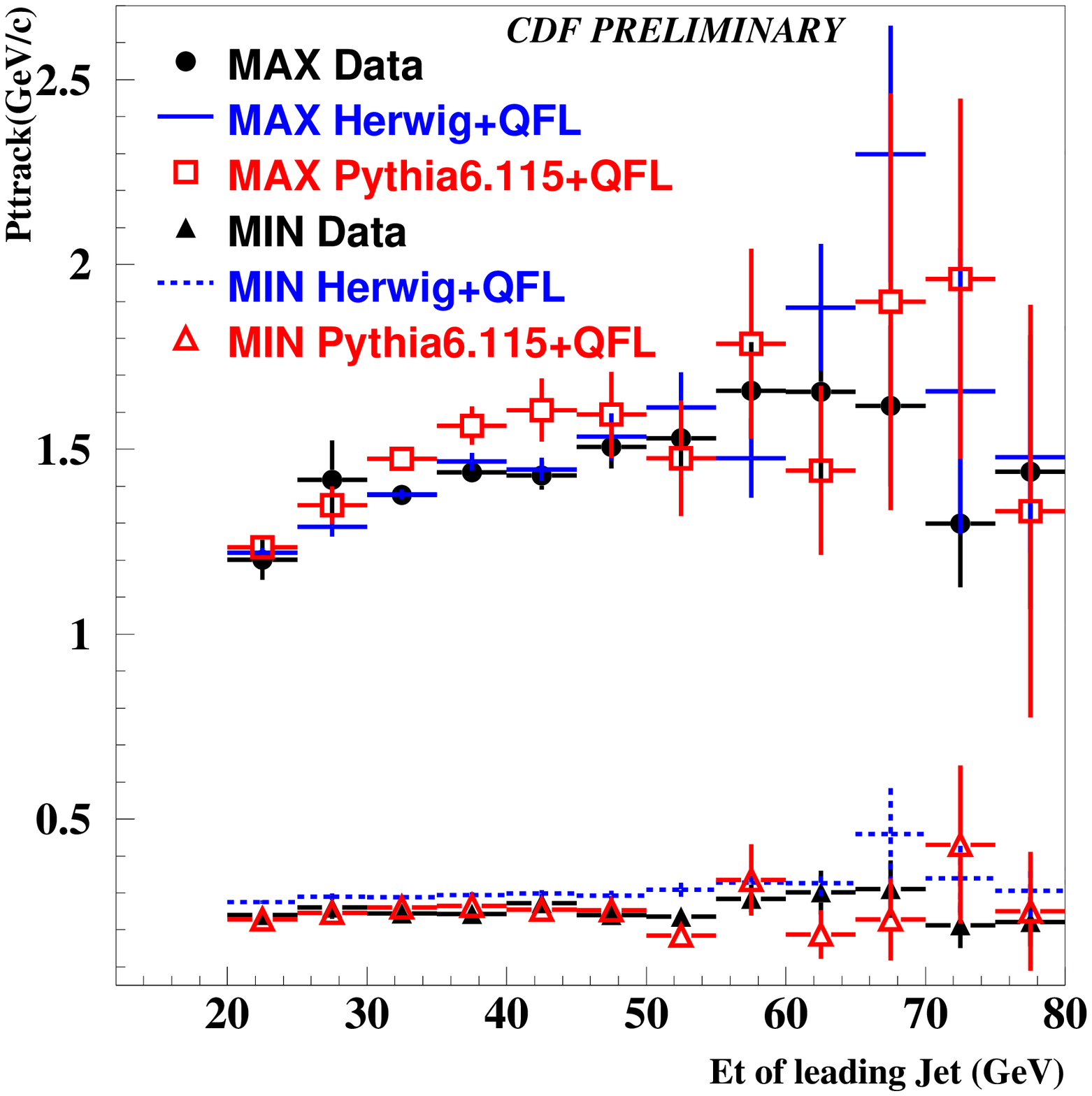}}
\caption{p$_{t_{cone}}$ inside the $max$ and $min$ cone as a function of the E$_t$  of the leading jet at $\sqrt{s}=1800$ GeV (on the left) and at $\sqrt{s}=630$ GeV (on the right). }
\label{Fig_track_max_min_tuned}
\end{figure}
In Figure~\ref{Fig_track_max_min_tuned} we plot the transverse momentum inside the $max$ and $min$ cone as a function of the E$_t$ of the leading jet at 1800 and 630 GeV.  
Both the Monte Carlo predictions and the data have a similar behavior for the $max$ and $min$ cone: the $min$ cone has a flat dependence on the E$_t$ of the leading jet while the $max$ cone increases as the leading jet E$_t$ increases. The behavior of the latter is due to the contribution of a third jet associated with the hard scatter.  However, one would expect also the $min$ cone energy to increase since the effects of higher order gluon radiation should become apparent for the higher leading jet energies.
Pythia has been tuned in order to reproduce the data, but the $max$ cone is still somewhat higher than the data for the larger leading jet E$_t$ region. The regularization scale of the transverse momentum spectrum for multiple interactions (p$_{t_{0}}$) depends on the center-of-mass energy and has been set to 2. and 1.4 GeV/c at 1800 and 630 GeV respectively.\par
\begin{figure}[h]
\centering
\mbox{
\hspace{-.5cm}
\includegraphics[height=7.8cm]{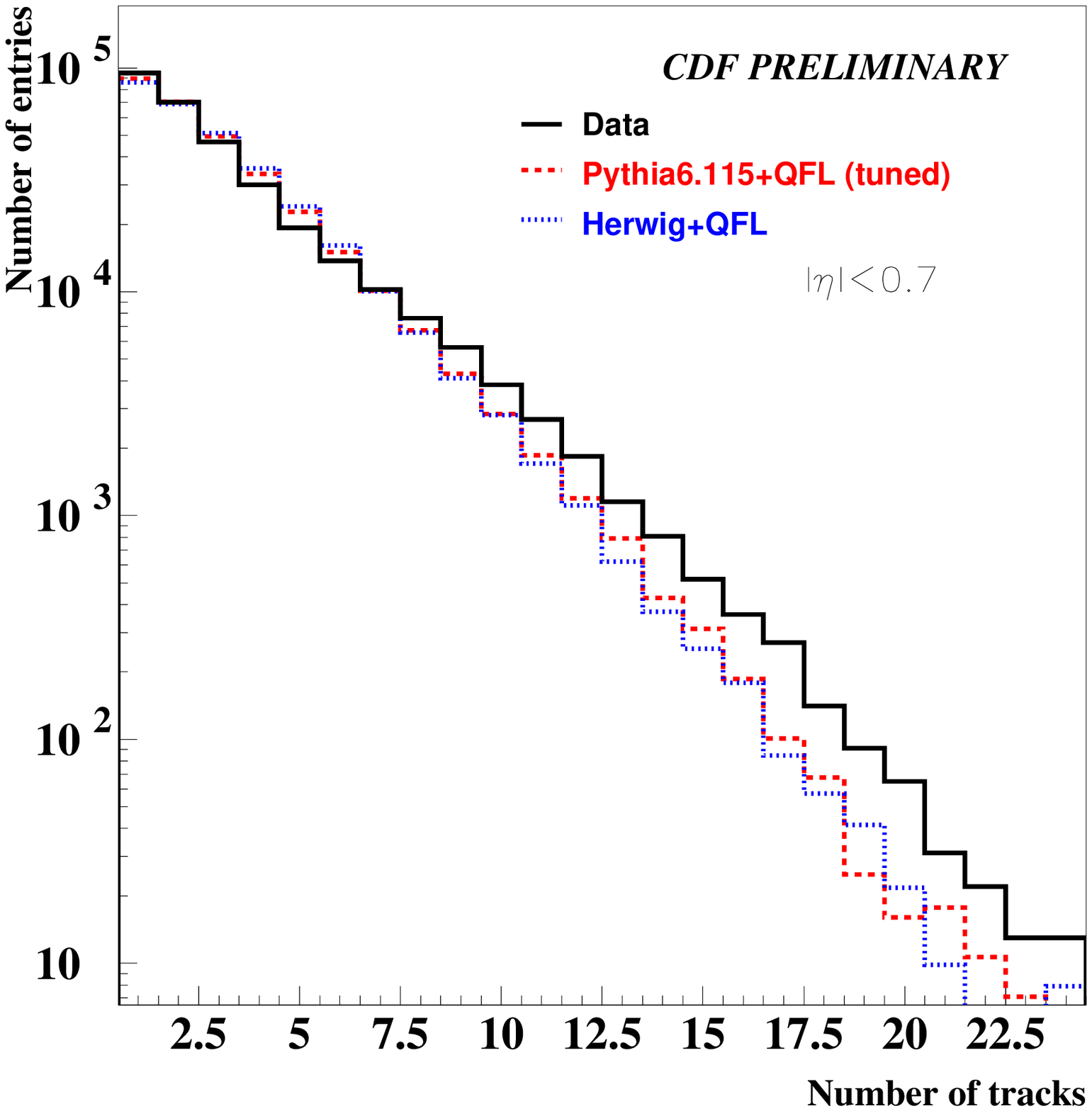}
\hspace{-0.5cm}
\includegraphics[height=7.8cm]{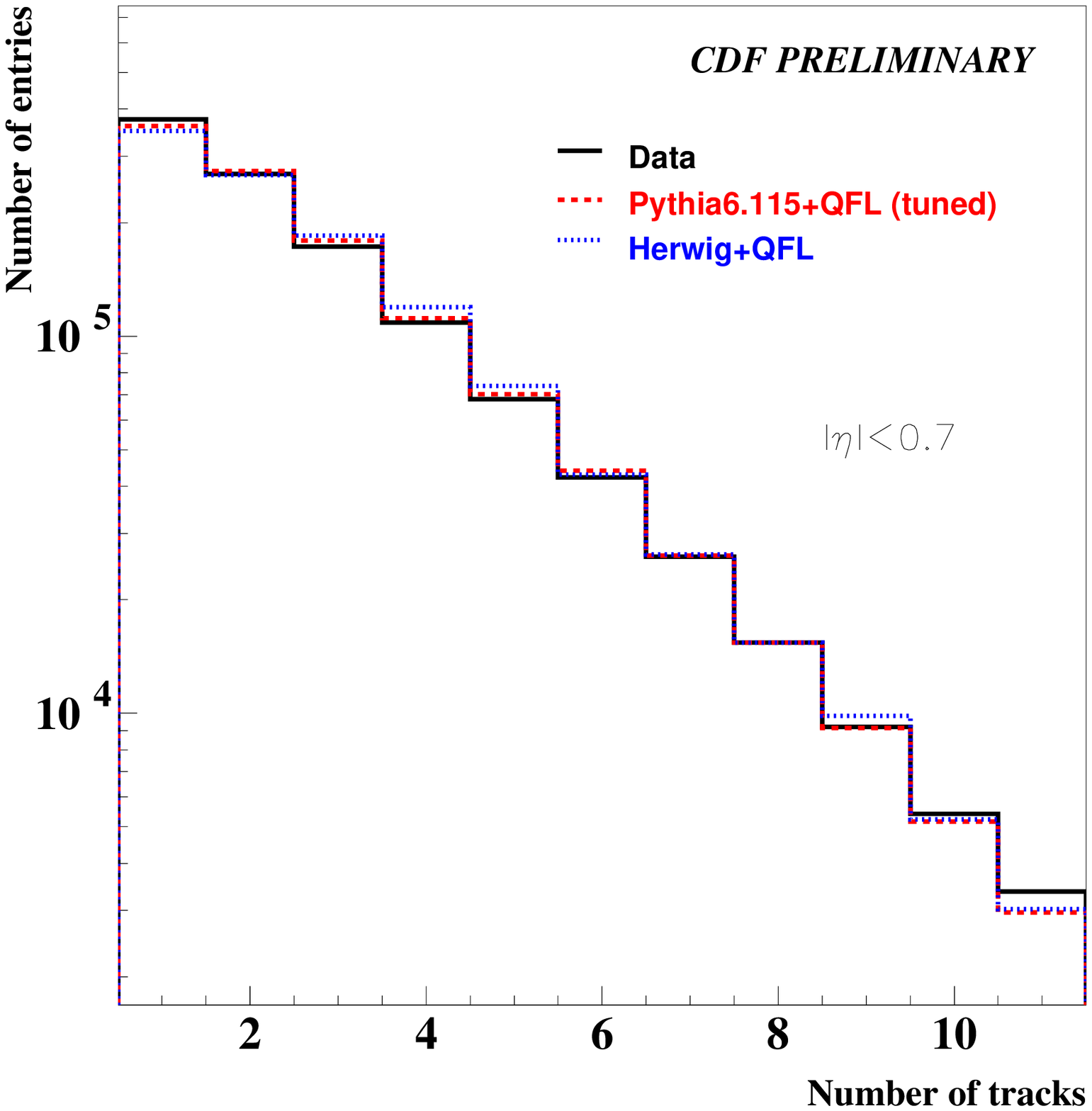}}
\caption{Multiplicity distribution in minimum bias events at $\sqrt{s}=1800$ GeV (left) and at $\sqrt{s}=630$ GeV (left).} 
\label{Fig_nt_her}
\end{figure}
\begin{figure}[h]
\centering
\mbox{
\hspace{-.5cm}
\includegraphics[height=7.8cm]{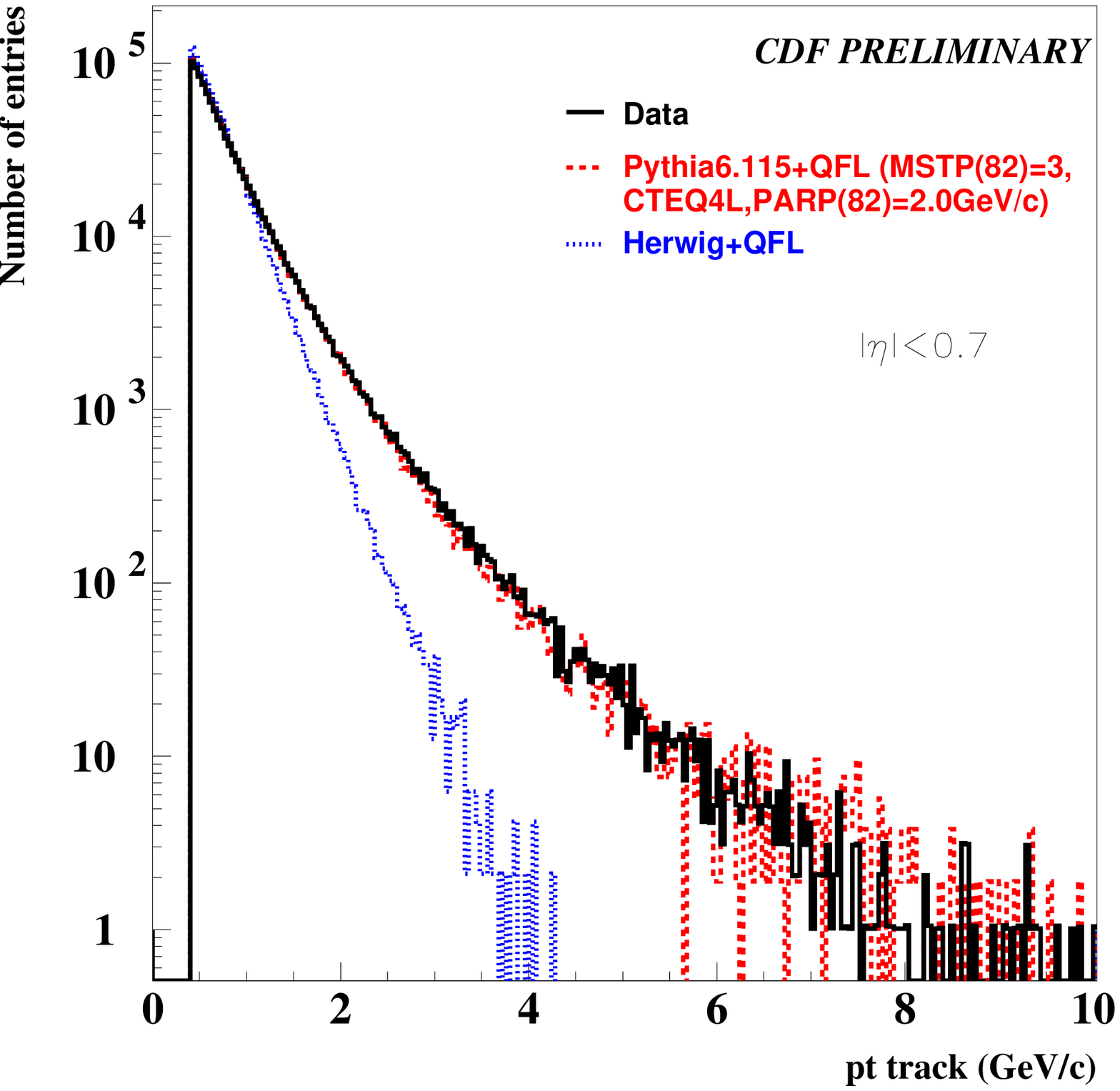}
\hspace{-0.5cm}
\includegraphics[height=7.8cm]{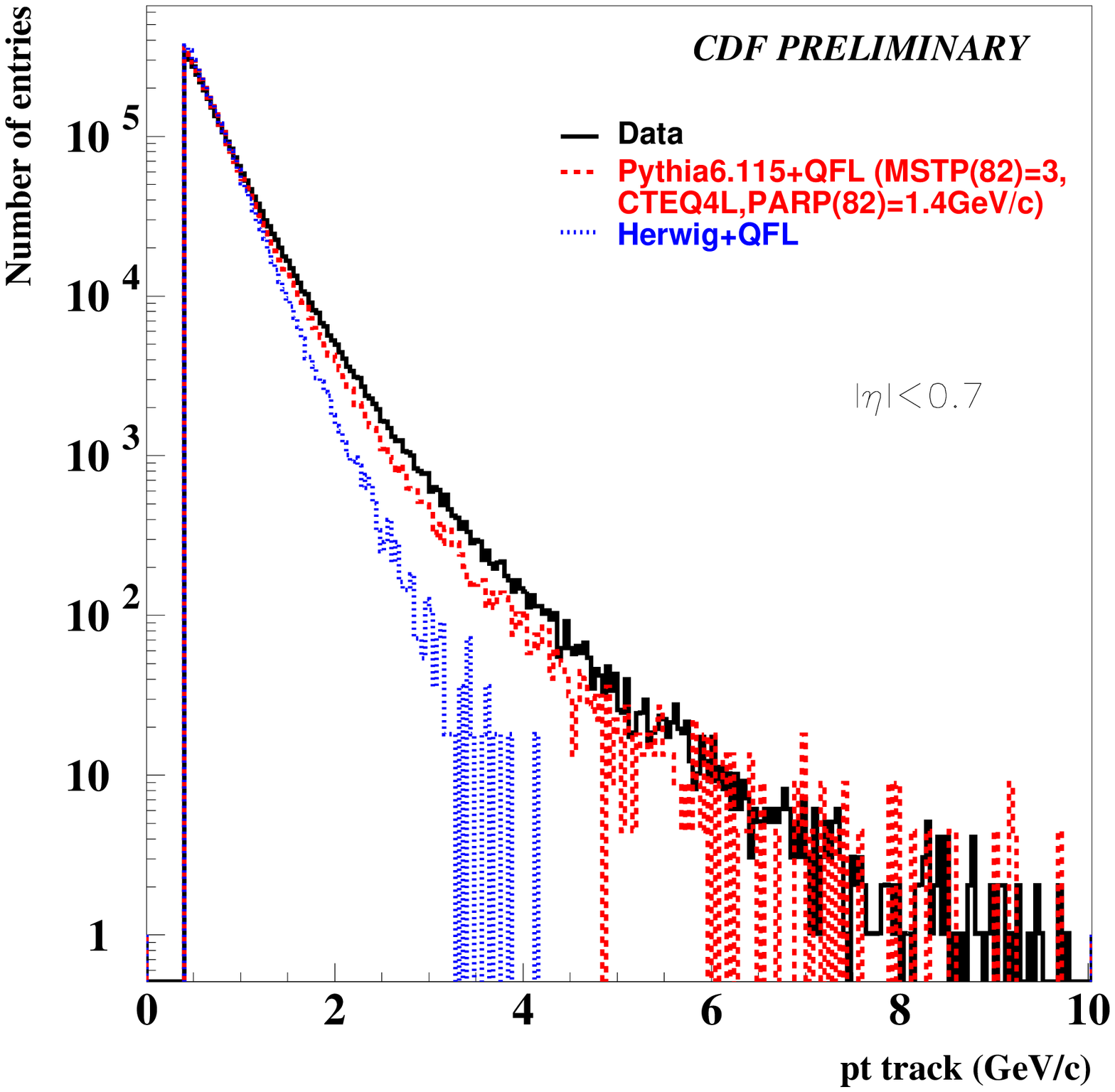}}
\caption{Distribution of the transverse momentum of tracks in the minimum bias sample at $\sqrt{s}=1800$ GeV (left) and at $\sqrt{s}=630$ GeV (right) in logarithmic scale.}
\label{Fig_pt_her}
\end{figure}

We perform similar investigations on minimum bias events by choosing a random
cone of radius 0.7 in the central region $|\eta|<0.3$.
We observe that for the same available energy the underlying event in a $hard$ scattering is considerable more active than in a $soft$ collision. The transverse momentum inside a random cone in the central rapidity region of the CDF detector in minimum bias data varies up to $20\%$ from the $min$ cone in jet events, according to vertex selection criteria. 
Figure~\ref{Fig_nt_her} shows the track multiplicity in the region $|\eta|<0.7$ at both 1800 and 630 GeV, with the number of entries in the simulation normalized to the  number in the data.
At 630 GeV, both generators describe the multiplicity distribution well, while at 1800 GeV neither Herwig nor Pythia reproduce the high multiplicity end of the distribution. 
The transverse momentum distribution of the tracks, shown in Figure~\ref{Fig_pt_her}, is generally not well predicted
by Herwig, with no tracks with a transverse momentum larger than 4
GeV/c.
The absence of high p$_t$ tracks indicates the lack of a semi-hard physics
description in the Herwig model of minimum bias events. Pythia
describes the transverse momentum distribution considerably better, but still fails to generate enough high p$_t$ tracks.

\section{Conclusions}

The study of the transverse momentum in jet events showed  that  data, Herwig and Pythia exhibit a similar behavior. However,  none of the examined Monte Carlo programs is able to describe all the properties of the data. \par
An improved understanding of {\it soft} interactions at the Tevatron is desired for predictions at the LHC, where the reconstruction of jets will be a difficult  task. The ability  to detect additional low p$_t$ jets, due to minimum bias events, is an important tool for the reduction of the background in many physics channels. LHC has to deal not only with underlying event energy, but also with energy coming from minimum bias events which are superimposed to the hard event (pile-up). 
Since at LHC about 25 minimum bias events on average are superimposed to a hard event, Herwig and Pythia predict about 8-12 GeV/c in a jet of cone size 0.7, solely due to charged particles related to ambient energy.\par

\end{document}